\documentclass[prd,showpacs,twocolumn,nofootinbib]{revtex4}

\usepackage[centertags]{amsmath}
\usepackage{amsfonts}
\usepackage{amssymb}
\usepackage{epsfig}
\usepackage{hyperref}

\DeclareMathOperator{\pr}{pr}

\begin{document}

\title{\textbf{Radiation reaction for multipole moments}}

\date{\today}

\author{P.O. Kazinski}

\email{kpo@phys.tsu.ru}

\affiliation{Department of Physics, Tomsk State University, Tomsk,
634050 Russia}

\begin{abstract}

We propose a Poincar\'{e}-invariant description for the effective
dynamics of systems of charged particles by means of intrinsic
multipole moments. To achieve this goal we study the effective
dynamics of such systems within two frameworks -- the particle
itself and hydrodynamical one. We give a relativistic-invariant
definition for the intrinsic multipole moments both pointlike and
extended relativistic objects. Within the hydrodynamical framework
we suggest a covariant action functional for a perfect fluid with
pressure. In the case of a relativistic charged dust we prove the
equivalence of the particle approach to the hydrodynamical one to
the problem of radiation reaction for multipoles. As the
particular example of a general procedure we obtain the effective
model for a neutral system of charged particles with dipole
moment.

\end{abstract}


\maketitle

\section{Introduction}

We address this paper to the problem of the construction in the
classical electrodynamics framework of an effective model
describing in some approximation the dynamics of an extended
object consisting of charged particles. The word effective means
that we take into account the self-interaction or radiation
reaction force exerting on such a system and thereby eliminate
from dynamical equations the electromagnetic field produced by
charged particles.

The simplest models of this kind are the model of a point charged
particle obeying the Lorentz-Dirac equations \cite{Lor,Dir} and
its generalizations to curved or higher dimensional space-times
\cite{DeWBr,Kos,rrmp}. Other models, which can be regarded as ad
hoc effective models of this kind, emerge at the classical
description of charged spinning particles
\cite{Frenk,Corb,BMT,Rowe,Univers,Fryd}. In \cite{str} the
effective model is formulated and investigated for a high-current
beam of charged particles taking into account the leading
self-interaction correction. In the paper \cite{OrRo} the
effective equations of motion are obtained for the center of mass
of a rigid charged body of arbitrary shape at the pointlike limit.
As expected they are the Lorentz-Dirac equations. A
nonrelativistic approach ($1/c$ expansion) to the effective
dynamics of a system of charged particles considering the
radiation reaction due to electric (magnetic) dipole and
quadrupole radiation can be found in \cite{LandLifsh,Itoh}.

In the present paper we obtain a Poincar\'{e}-invariant effective
model for a system of charged point particles as the model of a
point particle with some internal degrees of freedom like the
intrinsic dipole moment, the intrinsic magnetic moment etc.
Briefly, the procedure is as follows. We solve the Maxwell
equations for arbitrary worldlines of constituent charged
particles. Then we find the mean electromagnetic field in a small
neighborhood of the system under consideration by averaging over
all its small-scale fluctuations\footnote{The averaging procedure
is realized as the regularization of self-force and analogous to
averaging procedures applicable in the mean field and
renormalization group methods (see, e.g., \cite{Ma}).} and
substitute this field into the expression for the Lorentz force.
Using the obtained system of equations we derive the equations of
motion for the center of mass of the object and the evolution of
intrinsic multipole moments. Neglecting terms of higher order of
smallness we break the chain of equations and arrive at the closed
system of equations of motion for the effective model. These steps
are scrutinized in Sections \ref{eq m mul} and \ref{eff dyn mul
mom}.

In Section \ref{eq m mul} we also give a Poincar\'{e}-invariant
definition of the intrinsic multipole moments. Besides, as the
particular case of a general construction the effective model for
a neutral pointlike system of charged particles is obtained at the
end of Section \ref{eff dyn mul mom}, where we estimate the energy
losses of such a system and investigate its free dynamics.

In Section \ref{hydr appr} we consider an alternative approach to
the description of systems of charged particles. Namely, we pass
from the particle framework to hydrodynamical one and suggest an
obvious generalization of the standard action functional for
particles to the hydrodynamical case. This action is found to
describe a relativistic dust. Then we generalize the definition of
intrinsic multipole moments to the hydrodynamical approach and
prove the equivalence of particle and hydrodynamical approaches to
the problem of radiation reaction for multipole moments.

In addition, in Section \ref{hydr appr} we generalize the action
functional for a relativistic dust to a relativistic perfect fluid
with pressure and define in a Poincar\'{e}-invariant manner the
intrinsic multipole moments for extended objects (charged fluids
or systems of charged particles) approximated by branes, i.e. we
give a relativistic definition of the linear density of dipole
moment for a string or the surface density of quadrupole moment
for a membrane etc. In concluding section we summarize the main
results and outline the prospects for further investigations.

\section{Equations of motion and multipoles}\label{eq m mul}

In this section we recall some basic formulas concerning classical
electrodynamics of many-particle systems and define multipoles for
such systems in a Poincar\'{e}-invariant manner.

Let $\mathbb{R}^{3,1}$ be $4$-dimensional Minkowski space with
coordinates $\{x^\mu\}$, $\mu=0,1,2,3$, and signature $(+,-,-,-)$.
In the space-time given a system of $N$ electrically charged point
particles with trajectories $x_a(\tau_a)$, $a=1,\ldots,N$. The
dynamics of the system in question are governed by the action
functional\footnote{Hereinafter we use the system of units in
which $c=1$. The Greek indices are raised and lowered using the
Minkowski metric $\eta_{\mu\nu}$ on $\mathbb{R}^{3,1}$. Square
(round) brackets at pair indices denote antisymmetrization
(symmetrization) without one half.}
\begin{equation}\label{action particl}
\begin{split}
    S[x_a(\tau_a),A(x)]&=\sum\limits_{a=1}^N{\left[-m_a\int{d\tau_a\sqrt{\dot{x}_a^2}}-\int{d^4xA_\mu
    j^\mu_a}\right]}\\
    &-\frac1{16\pi}\int{d^4xF_{\mu\nu}F^{\mu\nu}},
\end{split}
\end{equation}
where $F_{\mu\nu}=\partial_{[\mu}A_{\nu]}$ is the strength tensor
of the electromagnetic field, overdots denote derivatives with
respect to the parameter on the worldline and the electric
currents of point particles are
\begin{equation}\label{currents}
    j^\mu_a(x)=e_a\int{d\tau_a\delta^4(x-x_a(\tau_a))\dot{x}^\mu_a(\tau_a)}.
\end{equation}
The equations of motion for the action functional \eqref{action
particl} constitute the system of Maxwell-Lorentz equations and in
the Lorentz gauge $\partial^\mu A_\mu=0$ are given by
\begin{equation}\label{eqmotpart}
    m_a\frac{d}{d\tau_a}\left[\frac{\dot{x}_a^\mu}{\sqrt{\dot{x}^2_a}}\right]=e_aF^\mu_{\ \nu}(x_a)\dot{x}_a^\nu,\qquad \Box
    A^\mu=4\pi\sum\limits_{a=1}^N{j^\mu_a}.
\end{equation}

To obtain the effective model for a system of charged point
particles we, first of all, should solve the Maxwell equations for
arbitrary worldlines $x^\mu_a(\tau_a)$. The casual solution to the
Maxwell equations can be constructed by the use of retarded
Green's function $G(x)$ associated with the d'Alembert operator
\cite{Cour}:
\begin{equation}\label{green function}
    G(x)=\frac{\theta(x^0)}{2\pi}\delta(x^2).
\end{equation}
Thus the Li\'{e}nard-Wiechert potentials
\begin{equation}\label{Lien-Wiech}
\begin{split}
    A^\mu(x)&=2\sum\limits_{a=1}^Ne_a\int d\tau_a\theta(x^0-x^0_a(\tau_a))\\
    &\times\delta((x-x_a(\tau_a))^2)\dot{x}^\mu_a(\tau_a)
\end{split}
\end{equation}
give the solution to the Maxwell equations. They are interpreted
as the electromagnetic field created by the system of charged
particles. A solution of the homogeneous Maxwell equations, which
can be added to the Li\'{e}nard-Wiechert potentials, is regarded
as an external field.

For our purposes it is useful to parametrize the trajectories of
particles in the following way: let $z^\mu(\tau)$ be a naturally
parametrized worldline in Minkowski space, then we parametrize the
trajectories $x^\mu_a(\tau_a)$ by the parameter $\tau$ and claim
that
\begin{equation}\label{parametrization}
    \dot{z}_\rho(\tau)\xi^\rho_a(\tau)=0,\qquad \dot{z}^2(\tau)=1,
\end{equation}
where $\xi^\mu_a(\tau)=x^\mu_a(\tau)-z^\mu(\tau)$. The gauge
\eqref{parametrization} properly fixes parametrizations on the
worldlines under the condition
\begin{equation}\label{gauge fix cond}
    \dot{z}_\rho\dot{x}_a^\rho>0,
\end{equation}
which is obviously satisfied. The worldline $z^\mu(\tau)$ is
defined by the requirement
\begin{equation}\label{center of mass}
    \sum\limits_{a=1}^N{m_a\xi_a^\mu(\tau)}=0,
\end{equation}
and we call it as the center of mass.

By the intrinsic electric and magnetic multipole moments of the
system of point charged particles we understand irreducible
components of the tensors\footnote{Usually electric multipole
moments are defined as traceless tensors, since at large distances
in stationary case the electromagnetic field of a charged object
depends only on their traceless parts. But in the dynamical case
the whole tensors should be employed.}
\begin{equation}\label{multipoles}
    \sum\limits_{a=1}^N{e_a\xi^a_{\mu_1}\ldots\xi^a_{\mu_n}},\qquad \sum\limits_{a=1}^N{e_a\xi^a_{\mu_1}\ldots\xi^a_{[\mu_{n-1}}\dot{\xi}^a_{\rho}\pr^\rho_{\mu_n]}},
\end{equation}
respectively, where
$\pr_\mu^\nu=\delta^\nu_{\mu}-\dot{z}^\nu\dot{z}_{\mu}$.
Evidently, they are orthogonal to $\dot{z}^\mu$. For example, the
Li\'{e}nard-Wiechert potentials \eqref{Lien-Wiech} can be
rewritten in terms of the multipole moments as
\begin{equation}\label{mult decomp}
\begin{split}
    A_\mu(x)&=qD\dot{z}_\mu+D[d_\mu-n_\rho d^\rho\dot{z}_\mu]\\
    &+\frac12D[QDz_\mu+X^\rho Q_{\rho\sigma}X^\sigma
    D^2z_\mu+S_{\mu\nu}n^\nu\\
    &+(\dot{z}^\rho-n^\rho)\dot{Q}_{\rho\sigma}(\delta^\sigma_\mu-n^\sigma
    \dot{z}_\mu)]+O(\xi^3),
\end{split}
\end{equation}
where one must assign $\tau=\tau_{ret}$ after all differentiations. We have
introduced the notations
\begin{equation}\label{notations}
\begin{gathered}
    X_\mu=x_\mu-z_\mu(\tau),\qquad R=\dot{z}_\rho
    X^\rho,\qquad n_\mu=\frac{X_\mu}{R},\\
    D=\frac{d}{Rd\tau},\qquad \left.X^2\right|_{\tau=\tau_{ret}}=0,\qquad \left.X_0\right|_{\tau=\tau_{ret}}>0,
\end{gathered}
\end{equation}
and the multipole moments
\begin{equation}\label{q d S Q}
\begin{gathered}
    q=\sum\limits_{a=1}^N{e_a},\qquad d_\mu=\sum\limits_{a=1}^N{e_a\xi^a_\mu},\\
    Q_{\mu\nu}=\sum\limits_{a=1}^N{e_a\xi^a_\mu\xi^a_\nu},\qquad S_{\mu\nu}=\sum\limits_{a=1}^N{e_a\xi^a_{[\mu}\dot{\xi}^a_\rho\pr^\rho_{\nu]}},
\end{gathered}
\end{equation}
which are the total charge, intrinsic dipole, quadrupole and
magnetic moments respectively. Besides $Q$ is the trace of the
quadrupole moment. Higher terms in the expansion \eqref{mult
decomp}, which we denote as $O(\xi^3)$, can be expressed in terms
of the multipole moments \eqref{multipoles} as well.

In particular, when all the multipole moments are constant and the system moves
freely we obtain a relativistic generalization of the well known expression
\cite{LandLifsh}:
\begin{equation}
    A_\mu(x)=\frac{q}{R}\dot{z}_\mu-\frac{n_\rho d^\rho}{R^2}\dot{z}_\mu+\frac{3n^\rho\bar Q_{\rho\sigma}n^\sigma\dot{z}_\mu
    +S_{\mu\nu}n^\nu}{2R^3}+O(\xi^3),
\end{equation}
where $\bar Q_{\mu\nu}=Q_{\mu\nu}-\frac13\pr_{\mu\nu}Q$.

\section{Effective dynamics of multipole moments}\label{eff dyn mul
mom}

In this section we substitute the Li\'{e}nard-Wiechert potentials
into the Lorentz force and regularize the resulting ill-defined
expression. After the regularization we obtain an infinite series
in a regularization parameter and $\xi^a_\mu$. To truncate this
series we impose constraints on characteristic scales of the
charged object under consideration. Thereby we derive the
equations of motion for the effective model associated with a
system of point charged particles. In particular, we obtain the
effective model for a neutral pointlike object with intrinsic
dipole moment, estimate the energy losses of such a system and
describe its free dynamics.

The effective equations of motion of charged particles at lower orders in
$\xi_\mu^a$ look like
\begin{equation}\label{eqmotion for ch p}
\begin{split}
    m_a\frac{d}{d\tau}\left[\dot{z}_\mu+\dot{\xi}^a_\rho\pr^\rho_\mu\right]&=e_aF^{rr}_{\mu\nu}(x_a(\tau))\dot{x}^\nu_a\\
    +e_aF_{\mu\nu}(&\dot{z}^\nu+\dot{\xi}^\nu_a)+e_a\partial_\rho F_{\mu\nu}\xi^\rho_a\dot{z}^\nu+\ldots,
\end{split}
\end{equation}
where $F_{\mu\nu}$ is the strength of the external electromagnetic
field taken at the point $z^\mu(\tau)$. As one can see the linear
order in $\xi$ of particle's momentum is orthogonal to
$\dot{z}^\mu$, that is the reason that we define the intrinsic
magnetic multipole moments as projected by $\pr^\mu_\nu$. The
field strength $F^{rr}_{\mu\nu}$ is constructed from the
potentials \eqref{Lien-Wiech} and is given by
\begin{equation}\label{field str rr}
\begin{split}
    F^{rr}_{\mu\nu}(x_a(\tau))&=4\sum\limits_{b=1}^Ne_b\int
    ds\theta(X^0+\xi^0_{ab})\\
    &\times\delta'((X+\xi^{ab})^2)
    (X_{[\mu}+\xi^{ab}_{[\mu})(\dot{z}_{\nu]}+\dot{\xi}^b_{\nu]}),
\end{split}
\end{equation}
where $X_\mu=z_\mu(\tau)-z_\mu(s)$ and $\xi^{ab}_\mu=\xi^a_\mu(\tau)-\xi^b_\mu(s)$.
Now we expand the integrand of \eqref{field str rr} in powers of $\xi$ and arrive at
\begin{multline}\label{series}
    \delta'(X^2)(X_{[\mu}+\xi^{ab}_{[\mu})\dot{x}^b_{\nu]}
    +\delta''(X^2)((\xi^{ab})^2+2X^\rho\xi^{ab}_\rho)\\
    \times(\xi^{ab}_{[\mu}+X_{[\mu})\dot{x}_{\nu]}^b
    +2\delta'''(X^2)(\xi^{ab}_\rho
    X^\rho)^2(\xi^{ab}_{[\mu}+X_{[\mu})\dot{x}_{\nu]}^b+\ldots
\end{multline}
Here dots denote neglible in our approximation terms (see below).

After integration every term in the series \eqref{series} gives
rise to infinities, because the $\delta$-function and its
derivatives are ill-defined at the vertex of light-cone (see,
e.g., \cite{GSh} and discussions in \cite{rrmp,rrmlp,siss}). To
cure this problem we have to regularize the integrals, i.e. to
represent them as sequences of converging integrals. We apply the
so-called ``point-splitting'' regularization of the
$\delta$-function and its derivatives which lies in substraction
of a positive number $\varepsilon$ (the regularization parameter)
from their arguments. Under this procedure the support of the
$\delta$-function transforms into a hyperboloid which is a smooth
manifold. That is why the integrals become converging.

In fact, the regularization of the $\delta$-function is the
regularization of the Green function \eqref{green function}.
Therefore the physical meaning of the regularization procedure
consists in averaging over all small-scale fluctuations of the
electromagnetic field produced by the system of charged particles
up to the scale $\varepsilon$. That is to say the regularization
parameter is a characteristic scale of fluctuations of the
electromagnetic field of the charged object.

\begin{figure}
\epsfig{file=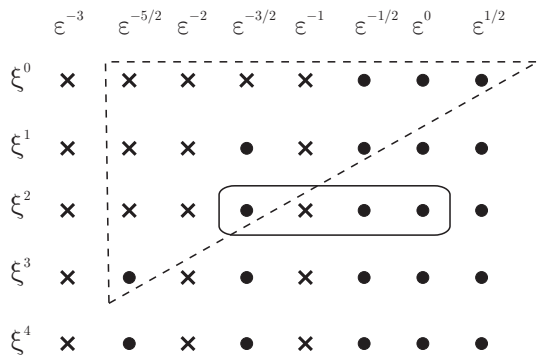,width=7cm} \caption{\label{contribut}
{\footnotesize The part of an infinite lattice depicting
contributions of the integral \eqref{field str rr}. The vanishing
terms are denoted by crosses. They are the terms at negative
integer powers of $\varepsilon$ and the terms at
$\xi^k\varepsilon^{-\frac{l+3}2}$, $k\leq l$. The dotted triangle
singles out the contributions, which we take into account for a
charged system meeting \eqref{scales relations}. Three marked dots
depict contributions of the self-force to the equations of motion
for the center of mass of a neutral pointlike object to the
accuracy of the first radiation correction. }}
\end{figure}

A useful mathematical framework for handling integrals with
integrands like \eqref{series} is elaborated in \cite{siss} and we
do not enlarge on it here. The structure of contributions of the
integral \eqref{field str rr} is as shown in Fig. \ref{contribut}.
All the terms at half-integer powers of $\varepsilon$ are
Lagrangian \cite{siss}, i.e. can be obtained by varying an
effective action, which in turn is obtained from the action
functional \eqref{action particl} by substituting the
Li\'{e}nard-Wiechert potentials \eqref{Lien-Wiech} in it and
applying the regularization procedure. The terms at integer powers
of $\varepsilon$ are responsible for radiation losses and not
restored by the effective action.

Let $l$ be a characteristic scale of variations of the trajectory
$z(\tau)$. Then, firstly, we average over\footnote{It is necessary
so as to the time-scale of variations of the fields
$\xi^a_\mu(\tau)$ would be larger than $\varepsilon^{1/2}$.
Otherwise the terms at higher powers of $\varepsilon$ make greater
contributions than at lower ones.} all oscillations of the fields
$\xi^a_\mu(\tau)$ the frequencies of which are greater than
$l^{-1}$. In that case the $n$-th derivative of $\xi^a_\mu(\tau)$
with respect to $\tau$ is of the order $\xi/l^n$, where $\xi$ is a
characteristic scale of fluctuations of the variables
$\xi^a_\mu(\tau)$. As the result for the integral \eqref{field str
rr} we have an infinite series in two dimensionless variables
$\xi/l$ and $\varepsilon^{1/2}/l$.

Secondly, one can see from Fig. \ref{contribut} the necessary
condition for the obtained asymptotic expansion makes sense is
\begin{equation}\label{scales relations 1}
    \xi\ll\varepsilon^{1/2}\ll l.
\end{equation}
Physically, the condition \eqref{scales relations 1} means that we
consider a charged system which is much smaller than the
characteristic scale of fluctuations of the electromagnetic field
(after averaging) while this characteristic scale is much smaller
than the characteristic scale of variations of the trajectory
$z(\tau)$.

Thirdly, in the lack of any a priori data on the system of charged
particles in question we make the assumption that there exists $n$
such that
\begin{equation}\label{scale relations on multipole}
    \overset{(n+k)}{M}\lesssim\overset{(n)}{M}(\xi/l)^k,
\end{equation}
for any positive integer $k$. Here $\overset{(n)}{M}$ denotes a
magnitude of the $n$-th multipole moment \eqref{multipoles}. In
other words we assume that beginning from the $n$-th order
multipole collective phenomena such as spontaneous magnetization
etc. are absent in the charged object, i.e. there is no any high
ordered structure in it. If that is the case then the leading
contribution to the self-force is contained in the first $n$
orders in $\xi$ of the asymptotic expansion.

In subsequent calculations we mostly deal with the case $n\leq2$, moreover we take
\begin{equation}\label{scales relations}
    \xi/l\ll(\varepsilon^{1/2}/l)^{\frac43},\qquad \varepsilon/l^2\lesssim\xi/l.
\end{equation}
These conditions mean that we consider a charged object which is much smaller than
$\varepsilon^{1/2}$, but is not so small to be pointlike. Some of the above
requirements could be relaxed, however it deserves individual investigations.

So, neglecting the terms of higher order of smallness we obtain
\begin{widetext}
\begin{multline}
    F^{rr}_{\mu\nu}(x_a(\tau))=-4\sum\limits_{b=1}^Ne_b\biggl\{\frac{3\varepsilon^{-\frac52}}8(\xi^{ba})^2\xi^{ba}_{[\mu}\dot{z}_{\nu]}
    +\frac{\varepsilon^{-\frac32}}{16}\left[4\xi^{ba}_{[\mu}\dot{x}^b_{\nu]}+(\xi^{ba})^2\ddot{z}_{[\mu}\dot{z}_{\nu]}-2(\ddot{z}^\rho\xi^{ba}_\rho+2\dot{z}^\rho\dot{\xi}^b_\rho)\xi^{ba}_{[\mu}\dot{z}_{\nu]}\right]\\
    +\frac{\varepsilon^{-\frac12}}{16}\left[(2-3\ddot{z}^\rho\xi^{ba}_\rho-6\dot{z}^\rho\dot{\xi}^b_\rho)\ddot{z}_{[\mu}\dot{z}_{\nu]}+2\ddot{z}_{[\mu}\dot{\xi}^b_{\nu]}+2\ddot{\xi}^b_{[\mu}\dot{z}_{\nu]}-2\xi^{ba}_{[\mu}\dddot{z}_{\nu]}-\frac{\ddot{z}^2}2\xi^{ba}_{[\mu}\dot{z}_{\nu]}\right]\\
    -\frac16\dddot{z}_{[\mu}\dot{z}_{\nu]}+\frac{\varepsilon^{\frac12}}{32}\left[3\overset{(4)}{z}_{[\mu}\dot{z}_{\nu]}+2\dddot{z}_{[\mu}\ddot{z}_{\nu]}+\frac52\ddot{z}^2\ddot{z}_{[\mu}\dot{z}_{\nu]}\right]
    \biggr\},
\end{multline}
where $\xi^{ba}_\mu=\xi^b_\mu(\tau)-\xi^a_\mu(\tau)$. Hence in our approximation the
average radiation reaction force acting on the charged point particle $a$ of the
system at issue reads
\begin{multline}\label{frr}
    e_aF^{rr}_{\mu\nu}(x_a(\tau))\dot{x}_a^\nu=
    -\sum\limits_{b=1}^Ne_ae_b\biggl\{\frac{3\varepsilon^{-\frac52}}2(\xi^{ba})^2\xi^{ba}_{\mu}
    +\frac{\varepsilon^{-\frac32}}{4}\left[(4+2\dot{z}^\rho(\dot{\xi}^b_\rho+\dot{\xi}^a_\rho))\xi^{ba}_{\mu}+(\xi^{ba})^2\ddot{z}_{\mu}-4\dot{\xi}^{a\rho}\xi^{ba}_\rho\dot{z}_\mu\right]\\
    +\frac{\varepsilon^{-\frac12}}{4}\left[(2-\dot{z}^\rho(\dot{\xi}^b_\rho+\dot{\xi}^a_\rho))\ddot{z}_\mu-2(\ddot{z}^\rho\dot{\xi}^a_\rho+\dot{z}^\rho\ddot{\xi}^b_\rho)\dot{z}_\mu+2\ddot{\xi}^b_\mu+\frac32\ddot{z}^2\xi^{ba}_\mu\right]\\
    -\frac23[\dddot{z}_\mu+\ddot{z}^2\dot{z}_\mu]+\frac{3\varepsilon^{\frac12}}{8}\left[\overset{(4)}{z}_\mu+\frac32\ddot{z}^2\ddot{z}_\mu+3\ddot{z}^\rho\dddot{z}_\rho\dot{z}_\mu\right]
    \biggr\}.
\end{multline}
\end{widetext}

Numerical coefficients at powers of the regularization parameter
depend on the regularization scheme applied. Different powers of
the regularization parameter should be regarded as independent
constants to be taken from an experiment. The averaging procedure,
which we have used, gives only the relations between orders of
magnitudes of these constants.

It is useful to introduce physical masses of particles
\begin{equation}\label{physical mass}
    \tilde{m}_a=m_a+\frac{e_aq}{2}\varepsilon^{-\frac12},
\end{equation}
and redefine $z(\tau)$ by the requirement \eqref{center of mass} with respect to
them. Then the equations of motion for the center of mass look like
\begin{equation}\label{eqmotion for center of mass}
\begin{split}
    (\tilde{M}+G)\ddot{z}_\mu&=F^{LD}_\mu+F^{(6)}_\mu
    +F_{\mu\nu}(q\dot{z}^\nu+\dot{d}^\nu)\\
    &+F_{\rho\sigma}\dot{d}^\rho\dot{z}^\sigma\dot{z}_\mu+d^\rho\partial_\rho
    F_{\mu\nu}\dot{z}^\nu,\\
    \dot{G}=F_{\rho\sigma}\dot{z}^\rho\dot{d}^\sigma,&\qquad \tilde{M}=\sum\limits_{a=1}^N{\tilde{m}_a},\qquad G=\frac{\varepsilon^{-\frac32}}2(qQ-d^2),
\end{split}
\end{equation}
where $F^{LD}_\mu=\frac23q^2[\dddot{z}_\mu+\ddot{z}^2\dot{z}_\mu]$ is the
Lorentz-Dirac force and
\begin{equation}\label{rigid term}
    F^{(6)}_\mu=-\frac{3\varepsilon^{\frac12}}{8}q^2\left[\overset{(4)}{z}_\mu+\frac32\ddot{z}^2\ddot{z}_\mu+3\ddot{z}^\rho\dddot{z}_\rho\dot{z}_\mu\right].
\end{equation}
Formally, the rigid term \eqref{rigid term} is nothing but the
counter term which, in addition to an ordinary mass
renormalization, compensates the divergencies emerging in
six-dimensional classical electrodynamics of a point charged
particle \cite{rrmp,Kos}. Such rigid terms have long been known
for nonrelativistic models of extended charged particles in the
four-dimensional space-time (see, e.g., \cite{Frad} and references
therein).

From \eqref{physical mass} and the condition of positiveness of
the total mass $\sum_{a=1}^N{m_a}$ we recover
\begin{equation}
    q^2/2\tilde{M}<\varepsilon^{1/2}.
\end{equation}
Furthermore, in what follows we imply the severer estimation
\begin{equation}\label{estimation for perturbations}
    q^2/\tilde{M}\lesssim\xi,
\end{equation}
which nevertheless is the reasonable one.

In order to close the system \eqref{eqmotion for center of mass}
and obtain the effective model for a charged object we have to
know the time evolution of the dipole moment. To this end we make
use of Eqs. \eqref{eqmotion for ch p}, \eqref{frr} again. At lower
orders they are
\begin{equation}\label{eqmotion at lower ord}
\begin{split}
     \tilde{m}_a\frac{d}{d\tau}\left[\dot{z}_\mu+\dot{\xi}^a_\rho\pr^\rho_\mu\right]&=e_a\varepsilon^{-\frac32}(q\xi^a_\mu-d_\mu)\\
     +e_a&F_{\mu\nu}(\dot{z}^\nu+\dot{\xi}_a^\nu)+e_a\xi_a^\rho\partial_\rho
    F_{\mu\nu}\dot{z}^\nu.
\end{split}
\end{equation}
The equations of motion describe the dynamics of mechanical
moments therefore we should determine relations between
electromagnetic and mechanical moments. The simplest way to do it
is to pick out different species of particles.

We call particles to be of the same species iff they have an equal
ratio $\lambda=e/\tilde{m}$. Let the system be made of $K$ species
of charged particles with charge to mass ratios $\lambda_s$,
$s=1,\ldots,K$. Then summing in Eqs. \eqref{eqmotion at lower ord}
over particles of the same species we obtain
\begin{equation}\label{dipole evolution}
\begin{split}
    \frac{d}{d\tau}\left[\dot{d}^s_\mu+(q_s+\ddot{z}^\rho
    d^s_\rho)\dot{z}_\mu\right]&=\lambda_s\varepsilon^{-\frac32}(qd^s_\mu-q_sd_\mu)\\
    +\lambda_sF_{\mu\nu}(q_s&\dot{z}^\nu+\dot{d}_s^\nu)+\lambda_sd_s^\rho\partial_\rho
    F_{\mu\nu}\dot{z}^\nu,
\end{split}
\end{equation}
where $q_s$ and $d^s_\mu$ are the total charge and dipole moment
corresponding to the species $s$. Equations \eqref{eqmotion for
center of mass} and \eqref{dipole evolution} must be solved
perturbatively starting from the Lorentz equations
\begin{equation}
    \tilde{M}\ddot{z}_\mu=qF_{\mu\nu}\dot{z}^\nu,
\end{equation}
and taking into account the estimation \eqref{estimation for perturbations}. In
doing so we eliminate all the higher derivatives of $z^\mu(\tau)$.

While the total charge of the object under consideration is
sufficiently small, i.e. $q^2/\tilde{M}\ll\xi$, in substituting to
\eqref{dipole evolution} one can disregard the Lorentz-Dirac
force, $G$-term and the rigid term in the equations of motion for
the center of mass \eqref{eqmotion for center of mass} and rewrite
the equations \eqref{dipole evolution} as
\begin{multline}\label{dipole evolution 1}
    \ddot{d}^s_\mu+\frac{q}{\tilde{M}}\frac{d}{d\tau}(d_s^\rho F_{\rho\sigma}\dot{z}^\sigma)\dot{z}_\mu=\lambda_s\varepsilon^{-\frac32}(qd^s_\mu-q_sd_\mu)\\
    +\left[q_s\left(\lambda_s-\frac{q}{\tilde{M}}\right)-\frac{q^2}{\tilde{M}^2}d_s^\rho F_{\rho\sigma}\dot{z}^\rho\right]F_{\mu\nu}\dot{z}^\nu\\
    +F_{\mu\nu}\left(\lambda_s\dot{d}_s^\nu-\frac{q_s}{\tilde{M}}\dot{d}^\nu\right)+\left(\lambda_sd_s^\rho-\frac{q_s}{\tilde{M}}d^\rho\right)\partial_\rho
    F_{\mu\nu}\dot{z}^\nu.
\end{multline}
In the case of $q^2/\tilde{M}\approx\xi$ some additional terms
appear.

Thus we obtain a closed system of equations describing in our
approximation the evolution of a charged object. To put it in
another way, singled out essential degrees of freedom and averaged
over or neglected the rest of ones we derive the aforementioned
effective model.

The equations \eqref{eqmotion for center of mass} and
\eqref{dipole evolution}, \eqref{dipole evolution 1} can be
regarded as the starting point for further perturbation theory in
$\xi$ to take into account higher multipoles effects. For example,
these last inevitably appear when one needs to consider the
self-interaction corrections to \eqref{dipole evolution} to the
same accuracy as in Eqs. \eqref{eqmotion for center of mass}.
Consequently, it is worthwhile to obtain the time evolution for
the second order in $\xi$ multipoles. Similarly to the above
considerations from Eqs. \eqref{eqmotion at lower ord} we find at
lower orders
\begin{widetext}
\begin{multline}\label{quadr evol}
    \lambda_s\dot{T}^s_{\mu\nu}+\pr_{(\mu}^\rho\dot{d}^s_\rho\ddot{z}_{\nu)}=\lambda_s\biggl\{\varepsilon^{-\frac32}\left[q\left(\dot{Q}^s_{\mu\nu}-\dot{Q}^s_{(\mu\rho}\bar\pr^\rho_{\nu)}\right)-\pr^\rho_{(\mu}\dot{d}^s_{\rho}d_{\nu)}\right]-\lambda_sT^{s\rho}_{(\mu}F_{\rho\nu)}\\
    \shoveright{+\left[\pr_{(\mu}^\rho\dot{d}^s_{\rho}-\frac12\ddot{z}^\rho\left(\dot{Q}^s_{\rho\varkappa}\pr^\varkappa_{(\mu}+S^s_{\rho(\mu} \right)\right]F_{\nu)\sigma}\dot{z}^\sigma+\frac12\left(\dot{Q}^s_{(\mu\sigma}\pr^\sigma_\rho-\bar\pr_{(\mu}^\sigma\dot{Q}^s_{\sigma\rho}-S^s_{(\mu\rho}\right)\partial^\rho F_{\nu)\varkappa}\dot{z}^\varkappa
    \biggr\},}\\
    \shoveleft{\frac{d}{d\tau}\left[\dot{Q}^s_{\mu\nu}+Q^s_{(\mu\rho}\ddot{z}^\rho\dot{z}_{\nu)}\right]+d^s_{(\mu}\ddot{z}_{\nu)}+\frac12\dot{z}_{(\mu}\ddot{z}^\rho\left[\dot{Q}^s_{\rho\sigma}\pr^\sigma_{\nu)}+S^s_{\rho\nu)}\right]=\lambda_s\biggl\{\varepsilon^{-\frac32}\left[2qQ^s_{\mu\nu}-d^s_{(\mu}d_{\nu)}\right]}\\
    \shoveright{+2T^s_{\mu\nu}+d^s_{(\mu}F_{\nu)\rho}\dot{z}^\rho-\frac12\left[\pr_{(\mu}^\sigma\dot{Q}^s_{\sigma\rho}+\dot{Q}^s_{(\mu\sigma}\bar\pr^\sigma_\rho+S^s_{(\mu\rho}\right]F^\rho_{\ \nu)}+Q^s_{(\mu\rho}\partial^\rho F_{\nu)\sigma}\dot{z}^\sigma
    \biggr\},}\\
    \shoveleft{\dot{S}^s_{\mu\nu}+d^s_{[\mu}\ddot{z}_{\nu]}+\frac12\dot{z}_{[\mu}\ddot{z}^\rho\left[\dot{Q}^s_{\rho\sigma}\pr^\sigma_{\nu]}+S^s_{\rho\nu]}\right]=\lambda_s\biggl\{\varepsilon^{-\frac32}d_{[\mu}d^s_{\nu]}+d^s_{[\mu}F_{\nu]\rho}\dot{z}^\rho}\\
    -\frac12\left[\pr_{[\mu}^\sigma\dot{Q}^s_{\sigma\rho}+\dot{Q}^s_{[\mu\sigma}\bar\pr^\sigma_\rho+S^s_{[\mu\rho}\right]F^\rho_{\ \nu]}+Q^s_{[\mu\rho}\partial^\rho
    F_{\nu]\sigma}\dot{z}^\sigma\biggr\},
\end{multline}
\end{widetext}
where $\bar\pr^\nu_\mu=\delta^\nu_\mu-\pr^\nu_\mu$, and
$T^s_{\mu\nu}=\sum\tilde{m}_a\pr_\mu^\rho\pr_\nu^\sigma\dot{\xi}^a_\rho\dot{\xi}^a_\sigma$
can be interpreted as the intrinsic stress (momentum flux)
tensor\footnote{Strictly speaking, the stress tensor is
$\int{d\tau\delta^4(x-z(\tau))T_{\mu\nu}(\tau)}$.} corresponding
to the species $s$. As all the intrinsic quantities we have
defined it is orthogonal to $\dot{z}^\mu$. Again we arrive at the
closed system of equations of motion \eqref{eqmotion for center of
mass}, \eqref{dipole evolution}, \eqref{quadr evol} for the
effective model.

Provided that the charged object is composed of particles of one
species\footnote{This case is of importance since it can be
considered as the base for perturbation theory in the deviation of
the ratios $\lambda_i$ from their mean value.} the intrinsic
dipole moment vanishes due to \eqref{center of mass} and the
system of equations \eqref{quadr evol} looks like
\begin{widetext}
\begin{multline}\label{quadr evol one species}
    \lambda\dot{T}_{\mu\nu}=\lambda\Bigl\{q\varepsilon^{-\frac32}\left(\dot{Q}_{\mu\nu}-\dot{Q}_{(\mu\rho}\bar\pr^\rho_{\nu)}\right)-\lambda T^{\rho}_{(\mu}F_{\rho\nu)}\\
    \shoveright{-\frac12\ddot{z}^\rho\left(\dot{Q}_{\rho\varkappa}\pr^\varkappa_{(\mu}+S_{\rho(\mu} \right)F_{\nu)\sigma}\dot{z}^\sigma+\frac12\left(\dot{Q}_{(\mu\sigma}\pr^\sigma_\rho-\bar\pr_{(\mu}^\sigma\dot{Q}_{\sigma\rho}-S_{(\mu\rho}\right)\partial^\rho F_{\nu)\varkappa}\dot{z}^\varkappa
    \Bigr\},}\\
    \shoveleft{\frac{d}{d\tau}\left[\dot{Q}_{\mu\nu}+Q_{(\mu\rho}\ddot{z}^\rho\dot{z}_{\nu)}\right]+\frac12\dot{z}_{(\mu}\ddot{z}^\rho\left[\dot{Q}_{\rho\sigma}\pr^\sigma_{\nu)}+S_{\rho\nu)}\right]=\lambda\Bigl\{2q\varepsilon^{-\frac32}Q_{\mu\nu}}+2T_{\mu\nu}\\
    \shoveright{-\frac12\left[\pr_{(\mu}^\sigma\dot{Q}_{\sigma\rho}+\dot{Q}_{(\mu\sigma}\bar\pr^\sigma_\rho+S_{(\mu\rho}\right]F^\rho_{\ \nu)}+Q_{(\mu\rho}\partial^\rho F_{\nu)\sigma}\dot{z}^\sigma
    \Bigr\},}\\
    \dot{S}_{\mu\nu}+\frac12\dot{z}_{[\mu}\ddot{z}^\rho\left[\dot{Q}_{\rho\sigma}\pr^\sigma_{\nu]}+S_{\rho\nu]}\right]=-\frac\lambda2\left[\pr_{[\mu}^\sigma\dot{Q}_{\sigma\rho}+\dot{Q}_{[\mu\sigma}\bar\pr^\sigma_\rho+S_{[\mu\rho}\right]F^\rho_{\ \nu]}+\lambda Q_{[\mu\rho}\partial^\rho
    F_{\nu]\sigma}\dot{z}^\sigma.
\end{multline}
\end{widetext}
The equations of motion for the center of mass \eqref{eqmotion for
center of mass} are modified into
\begin{equation}\label{eqmotion for center of mass one spec}
\begin{split}
    (\tilde{M}+G)\ddot{z}_\mu=&F^{LD}_\mu+F^{(6)}_\mu
    +qF_{\mu\nu}\dot{z}^\nu+\frac12Q^{\rho\sigma}\partial_{\rho\sigma}F_{\mu\nu}\dot{z}^\nu\\
    &+\frac12(\pr^{\rho\sigma}\dot{Q}_\sigma^\nu+\dot{Q}^{\rho\sigma}\bar\pr_\sigma^\nu+S^{\rho\nu})\partial_\rho F_{\mu\nu},\\
    \dot{G}=&\frac12\dot{Q}^{\rho\nu}\partial_\rho F_{\mu\nu}\dot{z}^\mu+\frac12(S^{\rho\nu}-\bar\pr^\rho_\sigma\dot{Q}^{\sigma\nu})\dot{F}_{\rho\nu},
\end{split}
\end{equation}
where we add the first corrections in $\xi$ to the external force
acting on the charged object as a whole.

For instance, for the unform external field $F_{\mu\nu}=const$ and
constant quadrupole moment $Q_{\mu\nu}=const$ the last equation of
the system \eqref{quadr evol one species} is nothing but the
Bargmann-Michel-Telegdi equation
\cite{Frenk,Corb,BMT,Barut,Bagrov}. The rest of Eqs. \eqref{quadr
evol one species} are equivalent to
\begin{equation}
    \lambda T_{\mu\nu}=\frac\lambda4S_{(\mu\rho}F^\rho_{\
    \nu)}+\frac14\dot{z}_{(\mu}\ddot{z}^\rho S_{\rho\nu)}-\lambda
    q\varepsilon^{-\frac32}Q_{\mu\nu}.
\end{equation}
The trace of this equation gives rise to the virial theorem in our
approximation. Notice as the quadrupole moment is constant there
is an inertial frame in which $Q_{\mu\nu}$ becomes diagonal for
all times. The condition $Q_{\mu\nu}\dot{z}^\nu=0$ implies either
this is the comoving frame and $\ddot{z}_\mu=0$ or at least one of
the space-like diagonal elements is zero, i.e. it is sufficiently
small. If only one diagonal element is zero (the case of a
thin-plate object) then the vector of intrinsic magnetic moment is
parallel to the respective axis. If two or three elements vanish
then $S_{\mu\nu}=0$.

As we have already noted the terms in Eqs. \eqref{frr} and
\eqref{eqmotion for center of mass} at half-integer powers of
$\varepsilon$ are Lagrangian. In order to simplify the Lagrangian
we write down it in the natural parametrization and keep only the
terms which are at most linear in the transverse gauge
\eqref{parametrization}:
\begin{multline}\label{lagr}
    L=\sum\limits_{a,b=1}^Ne_ae_b\Bigl\{\frac{3\varepsilon^{-\frac52}}{16}(\xi^{ba})^4\\
    +\frac{\varepsilon^{-\frac32}}{4}[(\xi^{ba})^2(1+\dot{z}^\rho\dot{\xi}^b_\rho)-2\dot{\xi}^{b\rho}\xi^{ba}_\rho\dot{z}^\sigma\xi^{ba}_\sigma]\\
    +\frac{\varepsilon^{-\frac12}}{32}[16+16\dot{z}^\rho\dot{\xi}^a_\rho+8\dot{\xi}^{a\rho}\dot{\xi}^b_\rho
    +3(\xi^{ba})^2\ddot{z}^2-8\dot{z}^\rho\dot{\xi}^a_\rho\dot{z}^\sigma\dot{\xi}^b_\sigma\\
    +16\dot{z}^\rho\xi^{ba}_\rho\ddot{z}^\sigma\dot{\xi}^b_\sigma-2(\ddot{z}^\rho\xi^{ba}_\rho)^2]-\frac{3\varepsilon^{\frac12}}{16}[\ddot{z}^2-3\ddot{z}^2\dot{z}^\rho\dot{\xi}^a_\rho+2\ddot{z}^\rho\ddot{\xi}^a_\rho]
    \Bigr\}.
\end{multline}
Varying the effective action with the Lagrangian density
\eqref{lagr} with respect to $\xi^a_\mu(\tau)$ we arrive at Eqs.
\eqref{frr}, varying it with respect to $z^\mu(\tau)$ and dropping
out the terms of higher order in $\xi$ we obtaining Eqs.
\eqref{eqmotion for center of mass}. In both cases the
Lorentz-Dirac force are not of course reproduced.

To conclude this section we derive the effective equations of
motion for a neutral system of charged particles at the pointlike
limit, i.e. at $q=0$ and $\xi\rightarrow0$. Rigorously, we
consider the system for which
\begin{equation}\label{scale structure q zero}
    \xi/l\ll(\varepsilon^{\frac12}/l)^5,\qquad q^2l^2\ll d^2,
\end{equation}
where $d$ is a magnitude of the intrinsic dipole moment. We also
assume the estimation \eqref{scale relations on multipole} is
fulfilled for $n=1$.

In what follows we are interested in the equations of motion for
the center of mass of the object in question taking into account
the first correction due to radiation, i.e. to the accuracy of the
leading non-Lagrangian contribution like the Lorentz-Dirac force
in the case of charged particle. At Fig. \ref{contribut} the
required terms are pictured as three dots at the third line
corresponding to the order of $\xi^2$. In view of \eqref{scale
structure q zero} other contributions are neglible.

The action for two Lagrangian terms can be evidently deduced from
\eqref{lagr}. Its Lagrangian density is equal to
\begin{equation}\label{lagr q zero}
\begin{split}
    L_{q=0}=-\frac{\varepsilon^{-\frac32}}{2}d^2+&\frac{\varepsilon^{-\frac12}}{16}[4\dot{d}^2-3d^2\ddot{z}^2-4(\dot{z}^\rho\dot{d}_\rho)^2\\
    &-8\dot{z}^\rho d_\rho\ddot{z}^\sigma\dot{d}_\sigma+2(\ddot{z}^\rho
    d_\rho)^2].
\end{split}
\end{equation}
Thus we have to find the non-Lagrangian term only. Tedious
calculations analogous to the case of a charged system show up
that the non-Lagrangian part of the self-force exerting on the
system as a whole looks like
\begin{widetext}
\begin{multline}\label{frr q zero}
    F^{rr}_\mu=\frac{4}{15}d^2\overset{(5)}{z}_\mu+\frac43d^\rho\dot{d}_\rho\overset{(4)}{z}_\mu+\frac23\left[2d^\rho\ddot{d}_\rho+\dot{d}^2+2(\dot{z}^\rho\dot{d}_\rho)^2+\ddot{z}^2d^2\right]\dddot{z}_\mu\\
    +\frac23\left\{d^\rho\dddot{d}_\rho+\frac{d}{d\tau}\left[(\dot{z}^\rho\dot{d}_\rho)^2+\frac32\ddot{z}^2d^2\right]\right\}\ddot{z}_\mu
    -\frac23\biggl\{\dot{d}^\rho\dddot{d}_\rho-(\ddot{z}^\rho\dot{d}_\rho+\dot{z}^\rho\ddot{d}_\rho)^2
    -2\ddot{z}^\rho\dot{d}_\rho\dot{z}^\sigma\ddot{d}_\sigma\\
    +2\dot{z}^\rho\dot{d}_\rho(\dddot{z}^\sigma\dot{d}_\sigma-\dot{z}^\sigma\dddot{d}_\sigma)
    -\ddot{z}^2\left[d^\rho\ddot{d}_\rho+7(\dot{z}^\rho\dot{d}_\rho)^2\right]-5\ddot{z}^\rho\dddot{z}_\rho d^\sigma\dot{d}_\sigma
    +d^2\left[\frac25\dot{z}^\rho\overset{(5)}{z}_\rho-\ddot{z}^4\right]\biggr\}\dot{z}_\mu\\
    -\frac23\left[\ddot{z}^\rho\dddot{d}_\rho+\dddot{z}^\rho\ddot{d}_\rho+\overset{(4)}{z}{}^\rho\dot{d}_\rho+\frac15\overset{(5)}{z}{}^\rho
    d_\rho+\ddot{z}^2(\ddot{z}^\rho\dot{d}_\rho-\dot{z}^\rho\ddot{d}_\rho)\right]d_\mu\\
    -\frac23\left[\dddot{z}^\rho\dot{d}_\rho-3\ddot{z}^2\ddot{z}^\rho
    d_\rho\right]\dot{d}_\mu
    +\frac23\dddot{z}^\rho
    d_\rho\ddot{d}_\mu-\frac23\dot{z}^\rho\dot{d}_\rho\dddot{d}_\mu.
\end{multline}
\end{widetext}
Gathering all the terms together we arrive at the following
equations for the center of mass:
\begin{equation}\label{eqmotion for center of mass q zero}
    M\ddot{z}_\mu=F^{lagr}_\mu+F^{rr}_\mu+F_{\mu\nu}\dot{d}^\nu+d^\rho\partial_\rho
    F_{\mu\nu}\dot{z}^\nu,
\end{equation}
where $F^{lagr}_\mu$ denote Lagrangian self-forces coming from
\eqref{lagr q zero}. To the same accuracy the time evolution of
the intrinsic dipole moment obeys
\begin{widetext}
\begin{multline}\label{dipole evolution 1 q zero}
    \frac{d}{d\tau}\left[\dot{d}^s_\mu+(q_s+\ddot{z}^\rho
    d^s_\rho)\dot{z}_\mu\right]=-\lambda_sq_s\biggl\{\varepsilon^{-\frac32}d_\mu+\frac{\varepsilon^{-\frac12}}4\left(2\ddot{d}_\mu+\frac32\ddot{z}^2d_\mu-\dot{z}^\rho\dot{d}_\rho\ddot{z}_\mu-2\dot{z}^\rho\ddot{d}_\rho\dot{z}_\mu\right)\\
    +\frac23\left[\dot{z}^\rho\dot{d}_\rho\dddot{z}_\mu+(\ddot{z}^\rho\dot{d}_\rho+2\dot{z}^\rho\ddot{d}_\rho)\ddot{z}_\mu+(\dot{z}^\rho\dddot{d}_\rho+2\dot{z}^2\dot{z}^\rho\dot{d}_\rho)\dot{z}_\mu-\ddot{z}^\rho\dddot{z}_\rho
    d_\mu-\ddot{z}^2\dot{d}_\mu-\dddot{d}_\mu\right]\biggr\}\\
    +\lambda_s\left[F_{\mu\nu}(q_s\dot{z}^\nu+\dot{d}_s^\nu)+d_s^\rho\partial_\rho
    F_{\mu\nu}\dot{z}^\nu\right].
\end{multline}
\end{widetext}
The contributions at the first and third lines are simply obtained
from Eqs. \eqref{frr}, while the term at the second line is
derived ab initio.

So, we derive the equations of motion for the effective model for
a neutral pointlike object with dipole moment. Note that in a
similar manner the equations of motion can be derived for the
effective model for a neutral pointlike object with magnetic
moment and vanishing dipole moment.

It is reasonable to suppose that the energy of the
self-interaction is much smaller than the rest energy of particles
in the absence of interaction, i.e.
\begin{equation}
    d^2/\varepsilon^{3/2}\ll M.
\end{equation}
In that case all the higher derivatives of $z_\mu(\tau)$ are
perturbatively expressed in terms of the first ones by means of
equations \eqref{eqmotion for center of mass q zero}. As far as
$\dddot{d}_\mu$ are concerned the similar procedure can be applied
to Eqs. \eqref{dipole evolution 1 q zero} under the assumption
that $\lambda_iq_il^2/\varepsilon^{\frac32}$ are of the order of
unity or less.

Let us extract some physical information from the huge equations
\eqref{eqmotion for center of mass q zero}, \eqref{dipole
evolution 1 q zero}. First of all, as one can see from Eqs.
\eqref{frr q zero} the total radiated power at small accelerations
of the center of mass of the object is equal to
\begin{equation}
    \mathcal{P}_\mu=\frac23\dot{d}^\rho\dddot{d}_\rho\dot{z}_\mu,
\end{equation}
and coincides with the known nonrelativistic expression (see,
e.g., \cite{LandLifsh}) for a dipole radiation. Besides, as it
follows from Eqs. \eqref{frr q zero} in the nonrelativistic
limit\footnote{Recall that on recovering the velocity of light
every overdot contains $1/c$.} the total radiated power is
\begin{equation}
    \mathcal{P}_\mu=\left[\frac23\dot{d}^\rho\dddot{d}_\rho-\frac4{15}\frac{d}{d\tau}(2d^\rho\dddot{d}_\rho+\dot{d}^\rho\ddot{d}_\rho)\right]\dot{z}_\mu.
\end{equation}
Whence the average emitted energy per unit time in the lab frame
amounts to $\frac23\overline{\ddot{\mathbf{d}}\ddot{\mathbf{d}}}$.

In the absence of external fields and at zero acceleration of the
center of mass the equations of motion \eqref{eqmotion for center
of mass q zero}, \eqref{dipole evolution 1 q zero} are reduced
to\footnote{The second equation in \eqref{eqmotions free q zero}
resembles the equation of motion for a damped linear oscillator.
The existence of self-oscillations of charged distributions has
been known long ago (see, e.g., \cite{BoWe}).}
\begin{equation}\label{eqmotions free q zero}
\begin{split}
    \frac{d}{d\tau}\left[\frac{\varepsilon^{-\frac12}}4\dot{d}^2+\frac{\varepsilon^{-\frac32}}2d^2\right]&=\frac23\dot{d}^\rho\dddot{d}_\rho,\\
    \left(1+\frac{\alpha\varepsilon^{-\frac12}}2\right)\ddot{d}_\mu&=-\alpha\varepsilon^{-\frac32}d_\mu+\frac{2\alpha}3\dddot{d}_\mu,
\end{split}
\end{equation}
respectively. In order to the system of equations \eqref{eqmotions
free q zero} possesses nontrivial solutions with the
characteristic scale of variations more than or equal to $l$ the
positive quantity $\alpha=\sum_{s=1}^K{\lambda_sq_s}$ is supposed
to meet the condition
\begin{equation}\label{estimation for alpha}
    \alpha^2/\varepsilon\ll\varepsilon^{\frac12}/l\ll1.
\end{equation}
The system \eqref{eqmotions free q zero} can be satisfied by the
second equation only provided the additional requirement
\begin{equation}\label{requirement on d}
    \dot{d}^2=const
\end{equation}
is fulfilled.

By the general procedure we must solve Eqs. \eqref{eqmotions free
q zero} perturbatively, but in our case we can do it exactly and
make approximations directly in a general solution. The
characteristic numbers associated to the second equation in
\eqref{eqmotions free q zero} are rather complicated and we write
down them in an approximate form
\begin{equation}
\begin{gathered}
    \lambda_{1,2}\approx\pm
    i\omega-\gamma,\qquad \lambda_3\approx\frac{\omega^2}{2\gamma},\\
    \omega^2=\alpha'\varepsilon^{-\frac32},\qquad \gamma=\frac{\alpha'^2}{3}\varepsilon^{-\frac32},\qquad \alpha'=\frac{2\alpha}{2+\alpha\varepsilon^{-\frac12}},
\end{gathered}
\end{equation}
taking into account the estimation \eqref{estimation for alpha}.
The solution corresponding to the third characteristic number is
known as runaway solution. We should drop it out since it does not
meet the initial hypothesis concerning the characteristic scale of
variations of the fields $\xi^a_\mu(\tau)$, i.e. $\lambda_3\gg
l^{-1}$. By the assumption \eqref{estimation for alpha} the
damping factor $\gamma$ is small, $\gamma\ll l^{-1}$, that is the
reason that the solution of the second equation in the system
\eqref{eqmotions free q zero}
\begin{equation}
\begin{gathered}
    d_\mu(\tau)=[a_\mu\cos{(\omega\tau)}+b_\mu\sin{(\omega\tau)}]e^{-\gamma\tau},\\
    a^2=b^2,\qquad a_\rho
    b^\rho=0,
\end{gathered}
\end{equation}
approximately satisfies the requirement \eqref{requirement on d}.
This solution represents a freely moving slowly rotating dipole.

\section{Hydrodynamical approach}\label{hydr appr}

In this section we regard a Poincar\'{e}-invariant hydrodynamical
approach to the description of a system of charged particles
thought as a relativistic perfect fluid. We briefly reformulate
basic definitions previously introduced within the particle
framework and prove the equivalence of the hydrodynamical and
particle approaches to the problem of radiation reaction for
multipoles under some assumptions given below. In conclusion of
this section we give a generalization of the notion of multipole
moments to extended relativistic objects (branes).

A simple general covariant generalization of the action
\eqref{action particl} to the hydrodynamical case can be
constructed as follows. Suppose given a $3$-brane $N$ with
coordinates $\{\tau^i\}$, $i=0,1,2,3$, that is embedded by a
diffeomorphism $x^\mu(\tau)$ into Minkowski space
$\mathbb{R}^{3,1}$. If the brane $N$ is the space-time itself one
can think about such a diffeomorphism as a general coordinate
transformation in it. Let us introduce $\rho^i(\tau)$ and
$e^i(\tau)$ that are vector densities on the brane $N$ and
describe a mass flow and an electric current respectively. Then an
obvious generalization of the action \eqref{action particl} takes
the form of\footnote{The action functional for the coupling of
branes to (non)-Abelian gauge fields is suggested in \cite{BSS}.}
\begin{equation}\label{action hydr}
\begin{split}
    S[x(\tau),A(x)]&=-\int_N{d^4\tau\sqrt{\rho^i\rho^jh_{ij}}}\\
    &-\int_{\mathbb{R}^{3,1}}{d^4x\left[A_\mu
    j^\mu+\frac1{16\pi}F_{\mu\nu}F^{\mu\nu}\right]},
\end{split}
\end{equation}
where $h_{ij}=\partial_ix^\mu\partial_jx^\nu\eta_{\mu\nu}$ is the
induced metric on the brane $N$, it being flat inasmuch as the
space-time is flat. The vector densities $\rho^i(\tau)$ and
$e^i(\tau)$ are supposed to vanish at spatial infinity. The
electric current density on the target space (the space-time) is
\begin{equation}
    j^\mu(x)=\int_N{d^4\tau\delta^4(x-x(\tau))e^i(\tau)\partial_ix^\mu(\tau)}.
\end{equation}
Of course, the $\delta$-function can be integrated out resulting
in the Jacobian, but we leave it intact to keep an analogy with
\eqref{currents}. In other words the matter action in
\eqref{action hydr} defines the dynamics on the
infinite-dimensional group of diffeomorphisms\footnote{A
group-theoretical approach to hydrodynamics can be found, for
example, in \cite{ArnKh}, where the equations of motion of a fluid
are regarded as the equations of geodesics on the
infinite-dimensional group of diffeomorphisms generated by the
fluid flows, which is equipped with a right-invariant Riemannian
metric.} $x(\tau^0,\bar\tau)$ of the space-like hypersurfaces:
$\tau^0=const$ and its image in the space-time. Hereinafter we
denote $\bar\tau=\{\tau^1,\tau^2,\tau^3\}$.

An invariance of the action functional \eqref{action hydr} under
gauge transformations of the electromagnetic potentials implies
the charge conservation law
\begin{equation}\label{conservation law charge}
    \partial_ie^i=0.
\end{equation}
Then the equations of motion for the matter obtained from the
action \eqref{action hydr} read
\begin{equation}\label{eqmotions hydr}
    \partial_i\left[\frac{\rho^i\rho^j\partial_jx_\mu}{\sqrt{\rho^2}}\right]=F_{\mu\nu}e^i\partial_ix^\nu.
\end{equation}
The matter energy-momentum tensor is\footnote{Notice that the
matter action \eqref{action hydr} can be easily generalized to the
case of a perfect fluid with pressure (see for other Lagrangians
\cite{Taub,Schutz,Brown,HajKij} and references therein).
Introducing a scalar density $p(\tau)$ on the brane $N$ we define
the action functional as
\[
    S[x(\tau)]=-\int_{N}{d^4\tau\left[\sqrt{\rho^2}+p\ln{\sqrt{\frac{\rho^2}{-h}}}\right]},
\]
where $h$ is the determinant of the induced metric, the vector
density $\rho^i(\tau)$ is the internal energy flow and $p(\tau)$
characterizes the fluid pressure.}
\begin{equation}
    T^{\mu\nu}_{mat}=\int{d^4\tau\delta^4(x-x(\tau))\frac{\rho^i\rho^j}{\sqrt{\rho^2}}\partial_ix^\mu\partial_jx^\nu},
\end{equation}
that is the energy-momentum tensor for a relativistic dust (see,
e.g., \cite{LandLifsh hyd}).

In a certain sense the hydrodynamical approach is reduced to the
particle one under the assumption
\begin{equation}\label{charge to mass ratio}
    e^i(\tau)=\lambda(\tau)\rho^i(\tau),
\end{equation}
where $\lambda(\tau)$ is some scalar function on $N$ describing
the charge to mass ratio. If that is the case Eqs.
\eqref{eqmotions hydr} give rise to
\begin{equation}\label{mass conservation law}
    \partial_i\rho^i=0,\qquad \rho^i\partial_i\lambda=0,
\end{equation}
i.e. to the mass conservation law. Provided the requirement
\eqref{charge to mass ratio} is fulfilled  the equations of motion
\eqref{eqmotions hydr} possess a ``partial'' reparametrization
invariance, which implies orthogonality of the equations to
$\rho^i\partial_ix^\mu$.

Now bearing in mind the assumption \eqref{charge to mass ratio} we
proceed to multipoles. Let $z^\mu(\sigma)$ be a naturally
parametrized worldline in Minkowski space, then similarly to
\eqref{parametrization} we claim that
\begin{equation}\label{parametrization hyd}
    \dot{z}_\rho(\tau^0)\xi^\rho(\tau)=0,\qquad \dot{z}^2(\tau^0)=1,
\end{equation}
where $\xi^\mu(\tau)=x^\mu(\tau)-z^\mu(\tau^0)$ and overdots
denote the derivative with respect to $\tau^0$. The condition
\eqref{parametrization hyd} partially specifies a parametrization
on the brane $N$ and foliates it into a family of the space-like
hypersurfaces $\tau^0=const$. Thereby the condition
\eqref{parametrization hyd} fixes the gauge and spoils the
reparametrization invariance of the equations of motion. The
trajectory $z^\mu(\tau^0)$ of the center of mass is defined by
\begin{equation}
    \int{d\bar\tau\rho^0(\tau)\xi^\mu(\tau)}=0.
\end{equation}
The definition of multipoles \eqref{multipoles} transforms into
\begin{equation}\label{multipoles hyd}
\begin{gathered}
    \int{d\bar\tau e^0(\tau)\xi_{\mu_1}(\tau)\ldots\xi_{\mu_n}(\tau)},\\
    \int{d\bar\tau
    e^i(\tau)\xi_{\mu_1}(\tau)\ldots\xi_{[\mu_{n-1}}(\tau)\partial_i{\xi}_{\rho}(\tau)\pr^\rho_{\mu_n]}}.
\end{gathered}
\end{equation}

It is easy to see that the Li\'{e}nard-Wiechert potentials are
expressed in terms of the multipole moments \eqref{multipoles hyd}
and their derivatives. Indeed, expanding in powers of $\xi$ the
$\delta$-function in the formula \eqref{Lien-Wiech} adjusted to
the hydrodynamical case we obtain expressions containing
$\xi_\mu(\tau)$ in the following way
\begin{equation}
\begin{split}
    \int &d\bar\tau\delta^{(k)}((x-z(\tau^0))^2)\\
    &\times\xi_{\mu_1}(\tau)\ldots\xi_{\mu_{n-1}}(\tau)e^i(\tau)\partial_i(z_{\mu_n}(\tau^0)+\xi_{\mu_n}(\tau)),
\end{split}
\end{equation}
that are obviously expressed in terms of the multipole moments
\eqref{multipoles hyd}.

As to mechanical moments are concerned the equations of motion for
particles \eqref{eqmotpart} suggest a general corresponding rule
between mechanical moments in the particle approach and
hydrodynamical one
\begin{equation}
    \sum\limits_a\leftrightarrow\int
    d\bar\tau,\qquad m_a\leftrightarrow
    \rho^0,\qquad d/d\tau\leftrightarrow
    \rho^i/\rho^0\partial_i.
\end{equation}
In our case the ambiguity of this rule must be settled by the
prescription that, at first, masses $m_a$ are removed from under
the differentiations with respect to $\tau$. For instance, the
intrinsic stress tensor arising in \eqref{quadr evol} is rewritten
as
\begin{equation}
    T_{\mu\nu}=\int{d\bar\tau\frac{\rho^i\rho^j}{\rho^0}\partial_i\xi_\rho\partial_j\xi_\sigma\pr^\rho_\mu\pr^\sigma_\nu}.
\end{equation}

In the sequel we do not repeat calculations of the preceding
section, but we show that the resulting equations of motion of the
effective model are identical to ones derived in the particle
framework. Besides, in spite of this assertion the equations of
motion for the center of mass \eqref{eqmotion for center of mass}
were rederived by direct calculations.

Similarly to the considerations regarding the Li\'{e}nard-Wiechert
potentials it can be proved that, firstly, the equations of motion
for the center of mass are expressed in terms of the mechanical
moments and multipoles \eqref{multipoles hyd}; secondly, on
introducing species\footnote{We emphasize once more that
separation into species is the simplest means to obtain a system
of ordinary differential equations from the integro-differential
one reducing to the multipole moments \eqref{multipoles hyd} only.
For example, another interesting choice of the ratio $\lambda$ is
to assume it to be the kernel of a differential operator
compatible with \eqref{mass conservation law}.} the evolution
equations for the intrinsic species moments both mechanical and
electric (magnetic) are expressed in terms of the intrinsic
species moments. In the hydrodynamical framework the species are
domains on the hypersurfaces $\tau^0=const$ of the constant ratio
$\lambda(\tau^0,\bar\tau)$. The form of these domains depends on
$\tau^0$ to be consistent with \eqref{mass conservation law}.

Whereas the equations of motion and expressions for multipoles in
the hydrodynamical case pass into respective expressions for the
particle case with the assumption
\begin{equation}
    \rho^i(\tau)=(\rho^0(\bar\tau),\mathbf{0}),
\end{equation}
we infer that the above-mentioned evolution equations are
identical to the equations derived in the preceding section.
Notice that the mass renormalization \eqref{physical mass} in the
hydrodynamical framework looks like
\begin{equation}
    \tilde{\rho}^i=\rho^i+\frac{e^iq}{2}\varepsilon^{-\frac12}\;\;\Rightarrow\;\;\partial_i\tilde\rho^i=0.
\end{equation}

In conclusion we point out that the elaborated approach to the
definition of the multipole moments is a quite general one and can
be used to describe the multipole moments for a charged fluid or
system of charged particles  approximated by not only a point
particle but also a brane, i.e. to define in a
Poincar\'{e}-invariant way the intrinsic linear density of dipole
moment for a string or the intrinsic surface density of magnetic
moment for a membrane etc.

Let us split the coordinates $\{\tau^i\}$ on the brane $N$ into
$\{\tau^a,\bar\tau^{\bar a}\}$, $\tau^0$ is included to
$\{\tau^a\}$. Suppose given a brane $M$ with coordinates
$\{\sigma^a\}$ that is embedded into the space-time by a smooth
mapping $z^\mu(\sigma)$. The induced metric on the brane $M$ we
denote by $h_{ab}$. Suppose the brane $N$ is also submersed into
$M$ by the mapping
\begin{equation}
    \varphi: N\rightarrow
    M,\qquad \sigma^a=\varphi^a(\tau,\bar\tau)=\tau^a,
\end{equation}
i.e. the mapping $\varphi$ matches systems of coordinates on the
branes $N$ and $M$. For our choice it merely identifies the
coordinates $\{\tau^a\}$ and $\{\sigma^a\}$. On the one hand this
mapping foliates the brane $N$ into a family of the space-like
surfaces $\tau^a=const$. On the other hand it foliates the brane
$N$ and, consequently, the space-time into a family of surfaces
diffeomorphic to the brane $M$. Then we claim that the vector
field $(\rho^b\rho^ch_{bc})^{-\frac12}\rho^a\partial_a$ is
projectable onto leaves of the latter foliation, i.e.
\begin{equation}
    [\partial_{\bar a},\frac{\rho^b}{\sqrt{\rho^2}}\partial_b]=0,
\end{equation}
hereinafter $\rho^2(\tau)=\rho^b(\tau)\rho^c(\tau)h_{bc}(\sigma)$.
Evidently this requirement can be satisfied by appropriately
chosen $\rho^i(\tau)$ and $x^\mu(\tau)$ for any given vector
density
\begin{equation}
    V^\mu(x)=\int_N{d^4\tau\delta^4(x-x(\tau))\rho^i(\tau)\partial_ix^\mu(\tau)},
\end{equation}
on the space-time.

The transverse condition \eqref{parametrization hyd} becomes
\begin{equation}
    \partial_az_\rho(\sigma)\xi^\rho(\tau)=0,
\end{equation}
where $\xi^\mu(\tau)=x^\mu(\sigma,\bar\tau)-z^\mu(\sigma)$. This
condition partially fixes a system of coordinates on the brane
$N$, i.e. it specifies the space-like surfaces $\tau^a=const$ on
the brane. The center of mass condition modifies into
\begin{equation}
    \int{d\bar\tau\sqrt{\rho^2(\tau)}\xi^\mu(\tau)}=0\;\;\Leftrightarrow\;\;\int{d\bar\tau\rho^a(\tau)\xi^\mu(\tau)}=0,
\end{equation}
and the intrinsic electric and magnetic multipole moments are
defined by
\begin{equation}
\begin{gathered}
    \int{d\bar\tau e^a(\tau)\xi_{\mu_1}(\tau)\ldots\xi_{\mu_n}(\tau)},\\
    \frac{\rho^a}{\sqrt{\rho^2}}\int{d\bar\tau
    e^i(\tau)\xi_{\mu_1}(\tau)\ldots\xi_{[\mu_{n-1}}(\tau)\partial_i{\xi}_{\rho}(\tau)\pr^\rho_{\mu_n]}},
\end{gathered}
\end{equation}
where
$\pr_\mu^\nu=\delta^\nu_\mu-\rho^a\rho^b\partial_az^\nu\partial_bz_\mu/\rho^2$
and we assume the relation \eqref{charge to mass ratio} is
fulfilled. The mechanical moments are generalized in an obvious
manner. Clearly, the so defined multipoles both mechanical and
electric (magnetic) are orthogonal to $\partial_az^\mu$ and zeroth
multipoles are conserved
\begin{equation}
    \partial_a\int{d\bar\tau e^a(\tau)}=\partial_a\int{d\bar\tau
    \rho^a(\tau)}=0,
\end{equation}
as long as corresponding currents are conserved.

\section{Concluding remarks}

Let us summarize the main results of our research. We have
investigated the effective dynamics of a system of charged
particles within two approaches -- the particle itself and
hydrodynamical one. We have described the effective dynamics of
such an object by means of a system of ordinary differential
evolution equations for the intrinsic multipole moments and the
center of mass. We have proved the equivalence of two examined
approaches to the problem of radiation reaction for multipole
moments. In passing we have derived the effective model for a
neutral pointlike system of charged particles.

These results can be extended to several directions. We point out
some of them only. It would be interesting to study the effective
dynamics of a string or membrane with intrinsic higher multipole
moments like that is given in \cite{str} for an electrically
charged string. Another direction for further research could be in
a study of a generalization of the effective dynamics of a
high-current beam of charged particles to include the matter term
proposed in Section \ref{hydr appr}. Other points could be the
nonrelativistic effective dynamics of a neutral pointlike object
with dipole moment or may be the effective dynamics of a neutral
pointlike object with magnetic moment and vanishing electric
dipole moment subjected to some simplifying assumptions.

\begin{acknowledgments}

I am grateful to I.V. Gorbunov and A.A. Sharapov for carefully
reading the draft of this manuscript and for their constructive
criticism. This work was supported by the RFBR grant 06-02-17352
and the grant for Support of Russian Scientific Schools
SS-5103.2006.2. Author appreciates financial support from the
Dynasty Foundation and International Center for Fundamental
Physics in Moscow.

\end{acknowledgments}


\end{document}